\documentclass[final,numberedheadings]{aipproc}
\layoutstyle{6x9}

\begin{document}
\title{Nuclear and Particle Astrophysics\\ at CIPANP 2003}

\author{\underline{Edward A. Baltz}}{address={ISCAP, Columbia Astrophysics
Laboratory, MC 5247, 550 W 120th St., New York, NY 10027},
altaddress={KIPAC, SLAC, M/S 78, 2575 Sand Hill Rd., Menlo Park, CA 94025 (from
Sept.\ 2003)}}

\author{James Stone}{
address={Physics Department, Boston University, Room 255, 590 Commonwealth
Ave., Boston, MA 02215}}

\begin{abstract}
In the nuclear and particle astrophysics session of CIPANP 2003 we heard talks
on a number of topics, focused for the most part into four broad areas.  Here
we outline the discussions of the standard cosmological model, dark matter
searches, cosmic rays, and neutrino astrophysics.  The robustness of
theoretical and experimental programs in all of these areas is very
encouraging, and we expect to have many questions answered, and new ones asked,
in time for CIPANP 2006.
\end{abstract}

\maketitle

\section{The Standard Cosmology}
In recent months the WMAP satellite has provided the clearest picture of
temperature fluctuations in cosmic microwave background (CMB) to date
\cite{wmap}.  The power spectrum of these fluctuations is a sensitive function
of the cosmological parameters, such as the total energy density, matter
density, baryon density, Hubble constant and others.  The emerging model was
convincingly confirmed: the universe is spatially flat, consisting mostly of
``dark energy'' that behaves like a cosmological constant, about 23\% dark
matter, and about 4\% baryons.  The sum of the neutrino masses is limited to be
less than 0.7~eV, much more strict than the bound from terrestrial
laboratories.  The Compton scattering optical depth implies an early
reionization at redshift $\sim17$.  These results will improve, as the WMAP
team plans to take at least 4 years of data.

Several profound questions arise from these results.  To start, the nature of
the dark matter and the dark energy are completely unknown.  We understand the
4\% in baryons, but the other 96\% is a mystery.  Furthermore, there is an
indication of a deficit of power on the largest scales (in e.g.\ the quadrupole
and octupole moments), but the meaning of this unclear.  A finite universe is
one among many unlikely possibilities.

The SNAP satellite\cite{snap} has been proposed to study the nature of the dark
energy by measuring the Hubble diagram (redshift -- distance relation) using
type Ia supernovae (SNIa) as standard candles, as has been done from the
ground.  Ground based measurements of this kind have already shown that there
is a large density of dark energy.  The SNAP team hopes to study the equation
of state $w:\; (p=w\rho)$ of this material.  A satellite is required to
significantly extend the ground--based results, as a large number of SNIa with
redshifts $z>1$ is required: this tests the universe in its decelerating phase.
The wavelengths of interest are redshifted into the infrared, thus the
necessity of a space--based instrument.  Furthermore, the exact nature of the
progenitors of SNIa is unknown.  A large sample can be split into many
subsamples to study systematic effects.  SNAP is fundamentally a large survey
telescope, and should have a broad science reach.

Big Bang Nucleosynthesis (BBN)\cite{bbn} tries to explain the primordial
abundances of light elements, in particular the stable isotopes of hydrogen,
helium and lithium.  As a function of only the baryon to photon ratio, these
abundances can be calculated by tracking the network of nuclear reactions in
the hot big bang.  The primordial abundance of deuterium depends sensitively
on the one free parameter, thus deuterium measurements can provide an accurate
assessment of the cosmological baryon density.  Lyman-$\alpha$ clouds obscuring
distant quasars presumably consist of predominantly unprocessed gas, and thus
reflect primordial abundances.  With the right column depth of hydrogen (not so
small that deuterium is unobservable, and not so large so that the damping
wings of hydrogen cover the deuterium line), the primordial deuterium abundance
is measured.  At present, the implied baryon density is fully consistent with
the WMAP value, with comparable errors.  To go further, many more such systems
would be needed.  Suitable systems are currently found at the rate of one per
year, so for now progress on the baryon density will come from CMB
measurements.

Cosmological observations have the potential to probe Planck-scale
physics\cite{transplanck}.  Inflationary theories usually predict that
observable wavelengths (e.g.\ galaxy and cluster scales) originated during
inflation as sub-Planck fluctuations.  Thus, inflation can in principle probe
quantum theories of gravity such as superstring theory.  With not overly
optimistic assumptions, a 1\% modulation of the CMB fluctuations might be
produced by Planck-scale physics during inflation, detectable in the next
decade or two.

\section{Dark Matter Searches}

The Standard Cosmological Model requires that 23\% of the energy density in the
universe is some form of non--baryonic nearly collisionless clustering matter.
A new stable particle would fit the bill, as has been known for several
decades.  In this regard the dark matter problem is more tractable than the
dark energy problem --- dark matter ``looks like'' something we understand,
while dark energy is completely mysterious.

Two possible candidates for dark matter have survived numerous tests and remain
viable.  The first is the lightest superpartner in supersymmetric extensions to
the Standard Model, which is naturally stable, weakly interacting, and
electrically neutral.  The second is the axion, arising in a compelling
solution to the strong CP problem.

Weakly Interacting Massive Particles (WIMPs), such as those in supersymmetric
models can be detected in sensitive low--background experiments by their rare
scattering from atomic nuclei.  The nuclear recoil deposits energy, which in
principle is detectable.  Two such detectors currently running are CDMS and
DAMA.

CDMS\cite{cdms} uses germanium and silicon detectors.  These are sensitive to
both phonons and ionization.  The ratio of these two signals powerfully
discriminates against background, as nuclear recoils exhibit much lower
ionization that most backgrounds (electrons and gamma rays).  CDMS-I ran in a
shallow site at Stanford, and the final WIMP exclusion results are now
available.  CDMS-II is currently running in the Soudan mine, with new results
anticipated by the end of 2003.

The DAMA\cite{dama} detector uses NaI scintillators.  They do not have the
background rejection capabilities of CDMS, but instead rely on the annual
modulation of the WIMP signal: as the Earth orbits the Sun, its relative
velocity with respect to the WIMP ``wind'' is modulated by several tens of km
s$^{-1}$, leaving a rate modulation of a few percent.  In the 4-year data such
a modulation is seen, though the implied mass and cross section are nearly
ruled out by other experiments (CDMS, EDELWEISS, ZEPLIN).  Three more years of
data have been released since the conference, and the modulation signal is
strengthened.  Furthermore, the successor experiment LIBRA is being installed
now.

The future of WIMP searches requires that ton-scale detectors be constructed.
The XENON proposal\cite{xenon} to use a two-phase detector for both
scintillation light and ionization is a promising possibility for scalability
to a one ton target mass.  Background can again be rejected by the lower levels
of ionization from nuclear recoils relative to electronic processes.  Small
prototypes are currently being tested.  The construction of a 10 kg prototype
is well underway.  The goal is to build a 100 kg module; with this a ton-scale
detector could feasibly be built in the next decade.

Axions in the range micro- to milli-eV remain a viable dark matter candidate.
They require a vastly different experimental approach: conversion to microwave
photons in a magnetic field.  These photons are then detected with with what is
essentially a very sensitive radio receiver.  The axion dark matter experiment
ADMX is ongoing, already scanning deep in the allowed model range\cite{axion}.
Upgraded receivers using SQUID amplifiers are on the way.  As light
pseudoscalars, axions have a bounded parameter space, as their interactions are
essentially the same as neutral pions.  The lowest allowed axion-photon
couplings are within reach.

\section{Cosmic Rays}

Energetic cosmic ray nuclei are an important probe of high energy processes in
astrophysics\cite{CRtheory}.  Supernova blast waves are capable of accelerating
protons to energies of 10$^{15}$ eV, the ``knee'' in the spectrum where the
power law shifts to a steeper value.  Very puzzling are the ultra high energy
cosmic rays with energies in excess of 10$^{20}$ eV.  Various acceleration
mechanisms have been proposed, involving FR II galaxies, interacting galaxies,
jets in radio sources (with Lorentz factor 10), gamma ray bursts (with Lorentz
factor 300), and others.  Top-down models have also been considered: UHECRs may
arise in the decay chains of supermassive particles.  Whichever mechanism
produces UHECRs, they must originate cosmologically nearby, within roughly 100
Mpc.  The GZK cutoff operates for nuclei above 10$^{20}$ eV / A, where the
threshold for pion photoproduction on the CMB is exceeded.  UHE photons have a
similar cutoff at lower energy, at the pair production threshold.  The nature
of the observed events above 10$^{20}$ eV is unknown, and difficult to
determine experimentally.  Nuclei such as iron, protons, photons, and neutrinos
are all possibilities.

The experimental situation in UHECRs is quite promising.  The HiRes
experiment\cite{hires} consists of two air fluorescence detectors situated 12.6
km apart, allowing stereoscopic viewing of the air showers induced in the
atmosphere.  The dataset exhibits the GZK cutoff, though at low significance.
The data are in agreement with AGASA, the Japanese particle detector, and the
AGASA data does not exhibit the cutoff.  The relative calibration between air
fluorescence and particle detectors is uncertain; the FLASH
collaboration\cite{flash} at SLAC expects to measure this to better than 10\%
accuracy.  HiRes does not see the clustering visible in the AGASA data, though
again the experiments are consistent.  This of course dilutes the AGASA
evidence for clustering.  The collaboration expects three more years of data to
be taken, which may clear up the situation.

The plan for the Pierre Auger project\cite{auger} is to use both particle
detectors (with 100\% duty cycle) and air fluorescence detectors (with 10\%
duty cycle).  The project promises greatly enhanced statistics and
cross-correlation between the two methods.  Two sites are planned: one in
Argentina to be fully operational in 2004, and one in the northern hemisphere.
3000 events per year per site above 10$^{19}$ eV are expected, with 30 per year
per site above $10^{20}$ eV.  The comparison of the two detector types will
allow more accurate energy measurements, and furthermore the identification of
the primary will be more certain (e.g.\ proton vs.\ Fe).  At the southern site,
a mountain range will act as a neutrino converter for low altitude primaries,
so UHE neutrinos can be studied as well.

At energies below 1 TeV, the Alpha Magnetic Spectrometer (AMS)\cite{ams} will
measure the spectra of cosmic ray species with higher accuracy that has been
possible.  Of particular interest are several exotic possibilities, including
antinuclei, dark matter, and quark matter ``strangelets''.  AMS-01, which flew
on the space shuttle, placed limits on the antihelium to helium ratio of
$10^{-6}$.  AMS-02 will be installed on the international space station as
early as 2005.  With three years of data, the sensitivity to antihelium will be
improved to $10^{-9}$.  AMS is sensitive to dark matter since annihilations in
the galactic halo may produce anomalous levels of antiprotons and positrons at
low to moderate energies, though very low energy particles are difficult to
study because of the geomagnetic cutoff.  AMS will have some sensitivity to
photons in the sub-TeV range as well.  A final possibility is the search for
stable strangelets.  They would be easily identified by their anomalous charge
to mass ratios.  Overall, AMS is a high statistics cosmic ray experiment, and
will advance our knowledge considerably.

\section{Neutrino Astrophysics}

The advent of large neutrino telescopes is exciting for high energy
astrophysics.  AGNs, GRBs, microquasars, and dark matter annihilations are all
interesting possible sources of neutrinos at GeV energies and higher.

The AMANDA array\cite{amanda} at the south pole uses strings of
photomultipliers (PMTs) deep in the ice to detect the tracks from (primarily
muon) neutrinos.  At 200m in diameter and 400m tall, it detects about 4
neutrinos per day, primarily atmospheric.  From the 2000 dataset, no point
sources were detected, but the photon flux limit was reached (based on the
expected relationship between photon and neutrino fluxes from hadronic
processes).  With timing information, GRBs are within a single order of
magnitude of detectability.  Over the next decade, AMANDA will be upgraded to
IceCube, with 80 strings totaling 4800 PMTs, with an effective volume of a
cubic km.  Neutrino energies from 50 GeV to more than a PeV will be studied.
At high enough energies, the three neutrino flavors can be disentangled:
electron from the shower characteristics, and tau from the ``double bang''
signature of the the recoil track and subsequent decay.  IceCube will be
installed starting in 2004, with completion expected in 2010.

ANTARES\cite{antares} is a complementary neutrino telescope to be built at the
floor of the Mediterranean.  12 cables totaling 1000 PMTs will cover 0.1
km$^2$, spaced at intervals of 60m.  Seawater has a longer scattering length
but a shorter absorption length than ice.  Thus, the PMTs must be closer
together, but the angular resolution is superior: above 10 TeV, a resolution of
$0.2^\circ$ is expected.  Between AMANDA / IceCube and ANTARES, most of the sky
will be covered.  One string is currently running, and completion is planned
for 2005.  The design for a km$^3$ upgrade is also underway.

The Super Kamiokande\cite{superK} neutrino detector has also begun to do
extra-solar neutrino astronomy.  A search for the diffuse neutrino background
from supernovae has been performed, and interesting limits have already been
set.

\section{Future Outlook}

The future of nuclear and particle astrophysics looks bright.  We have a
Standard Cosmological Model, but we understand very little about its matter and
energy contents.  We have observed very high energy processes, beyond the reach
of terrestrial accelerators, but the results are puzzling.  Several programs of
extra-solar neutrino astronomy are underway.  The level of experimental
activity is very encouraging.  We fully expect many new and interesting results
to be reported at CIPANP 2006.

\begin{theacknowledgments}
We thank all of the speakers in the nuclear and particle astrophysics session
of CIPANP 2003 for such interesting and informative talks, and we thank the
organizers for putting on such a good conference.
\end{theacknowledgments}

\bibliographystyle{aipproc}   % if natbib is available

\end{document}